\documentclass[aps,rmp,reprint,superscriptaddress,longbibliography,twocolumn]{revtex4-2}
\usepackage[table,cmyk]{xcolor}
\usepackage{amsmath, amsthm,commath,braket,booktabs}
\usepackage{calc,graphicx,bm,tikz}
\usepackage[charter,cal=cmcal,sfscaled=false]{mathdesign}
\usepackage[percent]{overpic}
\usepackage{xr}
\usepackage{multirow}
\definecolor{iffsred}{cmyk}{0.12,0.94,0.87,0.34}
\definecolor{uestcblue}{cmyk}{0.99,0.78,0.16,0.03}

\usepackage{hyperref}

\hypersetup{
  pdftitle={Practical Advantage of Classical Communication in Entanglement Detection},
  pdfauthor={Xing, Lv, Zhang et al.},
  pdfstartview=Fit,
  pdfpagelayout=SinglePage,
  colorlinks,
  linkcolor=uestcblue,
  citecolor=uestcblue,
  urlcolor=iffsred}

\begin{document}

\title{Practical Advantage of Classical Communication in Entanglement Detection}

\author{Wen-Bo Xing}\thanks{These authors contributed equally to this work}
\affiliation{CAS Key Laboratory of Quantum Information, University of Science and Technology of China, Hefei 230026, China}
\affiliation{CAS Center For Excellence in Quantum Information and Quantum Physics, University of Science and Technology of China, Hefei, 230026, China}
\affiliation{Institute of Fundamental and Frontier Sciences, University of Electronic Science and Technology of China, 611731, Chengdu, China}
\affiliation{Key Laboratory of Quantum Physics and Photonic Quantum Information, Ministry of Education, University of Electronic Science and Technology of China, 611731, Chengdu, China}
\author{Min-Yu Lv}\thanks{These authors contributed equally to this work}
\affiliation{CAS Key Laboratory of Quantum Information, University of Science and Technology of China, Hefei 230026, China}
\affiliation{CAS Center For Excellence in Quantum Information and Quantum Physics, University of Science and Technology of China, Hefei, 230026, China}
\author{Lingxia Zhang}\thanks{These authors contributed equally to this work}
\affiliation{Institute of Fundamental and Frontier Sciences, University of Electronic Science and Technology of China, 611731, Chengdu, China}
\affiliation{Key Laboratory of Quantum Physics and Photonic Quantum Information, Ministry of Education, University of Electronic Science and Technology of China, 611731, Chengdu, China}
\author{Yu Guo}
\affiliation{CAS Key Laboratory of Quantum Information, University of Science and Technology of China, Hefei 230026, China}
\affiliation{CAS Center For Excellence in Quantum Information and Quantum Physics, University of Science and Technology of China, Hefei, 230026, China}
\author{Mirjam Weilenmann}
\affiliation{Inria, T\'{e}l\'{e}com - LTCI, Institut Polytechnique de Paris, 91120 Palaiseau, France}
\author{Zhaohui Wei}
\affiliation{Yau Mathematical Sciences Center, Tsinghua University, Beijing 100084, China}
\affiliation{Yanqi Lake Beijing Institute of Mathematical Sciences and Applications, Beijing 101407, China}
\author{Chuan-Feng Li}
\affiliation{CAS Key Laboratory of Quantum Information, University of Science and Technology of China, Hefei 230026, China}
\affiliation{CAS Center For Excellence in Quantum Information and Quantum Physics, University of Science and Technology of China, Hefei, 230026, China}
\affiliation{Hefei National Laboratory, University of Science and Technology of China, Hefei 230088, China}
\author{Guang-Can Guo}
\affiliation{CAS Key Laboratory of Quantum Information, University of Science and Technology of China, Hefei 230026, China}
\affiliation{CAS Center For Excellence in Quantum Information and Quantum Physics, University of Science and Technology of China, Hefei, 230026, China}
\affiliation{Hefei National Laboratory, University of Science and Technology of China, Hefei 230088, China}
\author{Xiao-Min Hu}\email{huxm@ustc.edu.cn}
\affiliation{CAS Key Laboratory of Quantum Information, University of Science and Technology of China, Hefei 230026, China}
\affiliation{CAS Center For Excellence in Quantum Information and Quantum Physics, University of Science and Technology of China, Hefei, 230026, China}
\affiliation{Hefei National Laboratory, University of Science and Technology of China, Hefei 230088, China}
\author{Bi-Heng Liu}\email{bhliu@ustc.edu.cn}
\affiliation{CAS Key Laboratory of Quantum Information, University of Science and Technology of China, Hefei 230026, China}
\affiliation{CAS Center For Excellence in Quantum Information and Quantum Physics, University of Science and Technology of China, Hefei, 230026, China}
\affiliation{Hefei National Laboratory, University of Science and Technology of China, Hefei 230088, China}
\author{Miguel Navascu\'{e}s}\email{miguel.navascues@oeaw.ac.at}
\affiliation{IQOQI-Vienna, Austrian Academy of Sciences, Boltzmanngasse 3, 1090 Wien, Austria}
\author{Zizhu Wang}\email{zizhu@uestc.edu.cn}
\affiliation{Institute of Fundamental and Frontier Sciences, University of Electronic Science and Technology of China, 611731, Chengdu, China}
\affiliation{Key Laboratory of Quantum Physics and Photonic Quantum Information, Ministry of Education, University of Electronic Science and Technology of China, 611731, Chengdu, China}

\begin{abstract}
Entanglement is the cornerstone of quantum communication, yet conventional detection relies solely on local measurements. In this work, we present a unified theoretical and experimental framework demonstrating that one-way local operations and classical communication (1-LOCC) can significantly outperform purely local measurements in detecting high-dimensional quantum entanglement. By casting the entanglement detection problem as a semidefinite program (SDP), we derive protocols that minimize false negatives at fixed false-positive rates. A variational generative machine-learning algorithm efficiently searches over high-dimensional parameter spaces, identifying states and measurement strategies that exhibit a clear 1-LOCC advantage. Experimentally, we realize a genuine event-ready protocol on a three-dimensional photonic entanglement source, employing fiber delays as short-lived quantum memories. We implement rapid, FPGA-based sampling of the optimized probabilistic instructions, allowing Bob’s measurement settings to adapt to Alice’s outcomes in real time. Our results validate the predicted 1-LOCC advantage in a realistic noisy setting and reduce the experimental trials needed to certify entanglement. These findings mark a step toward scalable, adaptive entanglement detection methods crucial for quantum networks and computing, paving the way for more efficient generation and verification of high-dimensional entangled states.
\end{abstract}
	
\maketitle

Quantum entanglement~\cite{entanglement_measure_06,entangreview09Horodecki,entangReview09Gühne} is a necessary ingredient for most quantum communication protocols, such as dense coding~\cite{DenseCode18,guo2019advances}, quantum teleportation~\cite{Teleportation93,hu2023progress} and device-independent quantum cryptography~\cite{Cryptography07}. The reliable distribution and detection of entanglement between separate labs is therefore key to exploit the possibilities of a future quantum internet. 

Detecting entanglement in an experimental system is a two-part process~\cite{MultiEntangle05,convexEntang15,HighEntang18,Multiphotonentang16}: first, one needs to gather experimental data; second, one needs to argue, on the basis of the data, that the underlying quantum state is, in effect, entangled. Traditionally, the first part of the process has been tackled through full state tomography~\cite{ExpTomo93}; the non-separability of the resulting state is then established through general mathematical tools, such as the Peres-Horodecki criterion~\cite{PPT96} or the Doherty-Parrilo-Spedalieri (DPS) hierarchy of semidefinite programs~\cite{DPS02}. 

Full state tomography of a bipartite quantum system of local dimension $d$ requires $O(d^4)$ independent preparations of the state, as well as $O(d^4)$ different measurement settings. Therefore, as $d$ grows, the time required for the detection of high dimensional entanglement soon becomes prohibitive. This observation motivated a different approach, based on entanglement witnesses~\cite{witness00}. Namely, in order to certify entanglement, rather than estimating the full density matrix of the state, it is enough to estimate the average of a certain bipartite Hermitian operator: the entanglement witness. This task typically requires fewer local measurements and state preparations~\cite{FewMeaWitness11}. It also raises the question of how to estimate the value of said witness efficiently, or, more generally, given access to such a minimal set of measurements, how to certify entanglement with confidence with a modicum of state preparations, i.e., how to construct the best witness for the task at hand. 

Typically, the value of an entanglement witness $W$ is determined through local operations (LO), i.e., entanglement detection protocols where the measurement settings for each subsystem are chosen independently. After many rounds, measurement settings and outcomes are shared and the value of $W$ can be estimated without the need of storing any quantum state. However, it turns out that access to an (even relatively short-lived) quantum memory can make entanglement detection more efficient, without changing the local measurements to be performed. Specifically, one can implement a 1-way Local Operations and Classical Communication (1-LOCC) protocol, where the choice of measurement on one subsystem depends on the setting and the outcome of the other. 

The implementation of event-ready LOCC protocols in high-dimensional systems presents significant technical challenges due to the stringent requirements for a high dimension quantum memory~\cite{dong2023highly} and high-speed real-time control for adaptive high-dimensional measurements~\cite{sund2023high}. These challenges are further exacerbated by the complexity of preparing and certifying high-dimensional quantum systems, where the number of measurement settings and classical communication steps grows quadratically with the dimension~\cite{wang2018multidimensional}. So far, numerous entanglement-based LOCC operations have been successfully implemented in the qubit domain for tasks such as distributed quantum computing~\cite{Carrera2024Computing,liu2024nonlocal}, entanglement extraction~\cite{Yamamoto2003Extraction}, entanglement swapping and distribution~\cite{Pompili2021Distribution}, teleportation~\cite{Jin2010Teleportation,Ma2012Teleportation,Krauter2013Teleportation,Takeda2013Teleportation,Pfaff2014Teleportation,Chou2018Teleportation,Hermans2022Teleportation}, quantum state discrimination~\cite{Zhang2023aDiscrimination} and entanglement certification~\cite{entangleVerLO18,ExpEntangleVerLO20}. However, the implementation of high-dimensional LOCC operations remains elusive.

Notably,~\cite{QuanPereGame21} has shown theoretically that for specific quantum states, some 1-LOCC entanglement detection protocols can achieve significantly higher statistical confidence compared to the best LO protocols, especially in the case of high-dimensional entangled states. In the context of a quantum internet, where efficient and robust entanglement detection is essential, 1-LOCC protocols are particularly advantageous as they reduce the need for repeated one-photon transmissions and enable faster calibration of experimental devices, paving the way for scalable quantum communication systems.

Experimentally demonstrating such an advantage highlights the interdependence of the theoretical and experimental sides of the problem. We first turn our experimental scenario into an optimization problem~\cite{QuanPereGame21,OptimDectLOCC21}. However, the inputs to this problem are constrained by experimental considerations such as what states can be prepared and what measurements are possible. We in turn take these experimental constraints into another optimization problem, which we solve by using a generative machine learning algorithm~\cite{VGON24}. The solution contains experimental instructions which can be directly used in state preparation and protocol execution. Probabilistic instructions for the protocols are loaded into a field programmable gate array (FPGA) chip, which controls electro-optic modulators (EOMs) to enable fast real-time control of measurements, essential ingredients for implementing classical communication. For entanglement detection in $3$-dimensional states, we verify that the probability for false negatives of the single-shot protocol is smaller than that of any LO protocol with the same probability of false positives. Repeated realizations of the single-shot protocol would thus certify entanglement with a $p$ value impossible to achieve through LO methods~\cite{Araujo_2020}. 

\section*{Conceptual Setup}
Our results are obtained in the following setup.
Consider the experimental scenario where Alice and Bob aim to certify the entanglement of a shared quantum state $\rho$ with a set of local measurements $\{A_x^a\}$ ($\{B_y^b\}$), where $x$ ($y$) are labels for measurement settings and $a$ ($b$) are labels for the corresponding outcomes. The entanglement status, $\gamma$, which can be $Y$ (yes) or $N$ (no), is computed as a random function of the labels $a,b,x,y$.

This random function, together with the probabilistic choice of measurements, are called the \textit{protocol}. It contains the probabilistic instructions for Alice and Bob and incorporates the flexibility of the experiment.  What Alice and Bob are allowed to do in each round of the experiment, and in what order, can be modeled through linear constraints on the random function. Nevertheless, the goal of the experiment is the same for all protocols: minimize the probability of errors when computing the final outcome $\gamma$. Errors manifest as either type-I (false positive) or type-II (false negative), signifying that $\gamma=Y$ when $\rho$ is separable and $\gamma=N$ when $\rho$ is entangled, respectively.

The experimental scenario, therefore, corresponds to finding the optimal solution of the following convex optimization problem~\cite{QuanPereGame21,OptimDectLOCC21}:
\begin{align}
		\min_{P(x,y,\gamma|a,b)} \quad & p_2  \nonumber\\
		\text{subject to }\quad       & {\rm tr}(M_N\rho)=p_2,     \nonumber\\
		                                 & p_1 \mathbb{I}-M_Y \in \mathcal{S}^*;\quad M_{Y}=\mathbb{I}-M_{N}      \nonumber\\
		                               & M_{N}=\sum_{a,b,x,y}P(x,y,\gamma=N|a,b)(A_{x}^{a} \otimes B_{y}^{b}) \nonumber\\
		                               & P(x,y,\gamma|a,b) \in \mathcal{P}^{\{LO,1-LOCC\}}\label{eq:sdp},
\end{align}
where $P(x,y,\gamma|a,b)$ gives the protocol and $p_1$, the probability of type-I errors, is fixed at a reasonable value to limit the likelihood of mistakenly classifying states in the separable set $\mathcal{S}^*$ as entangled.

Once the target state $\rho$ and available measurements $\{A_x^a,B_y^b\}$ are specified, problem (\ref{eq:sdp}) is a semi definite program (SDP), the solution is the minimal false-negative probability $p_2^*$. In addition, it also gives the probabilistic instructions $P(x,y,\gamma=N|a,b)^*$ on how the experiment must be carried out in each round, which allows Alice and Bob to cooperatively implement the POVM $M_{N}^*$.
\begin{figure}[tbph!]
    \centering
    \includegraphics[width=0.45\textwidth]{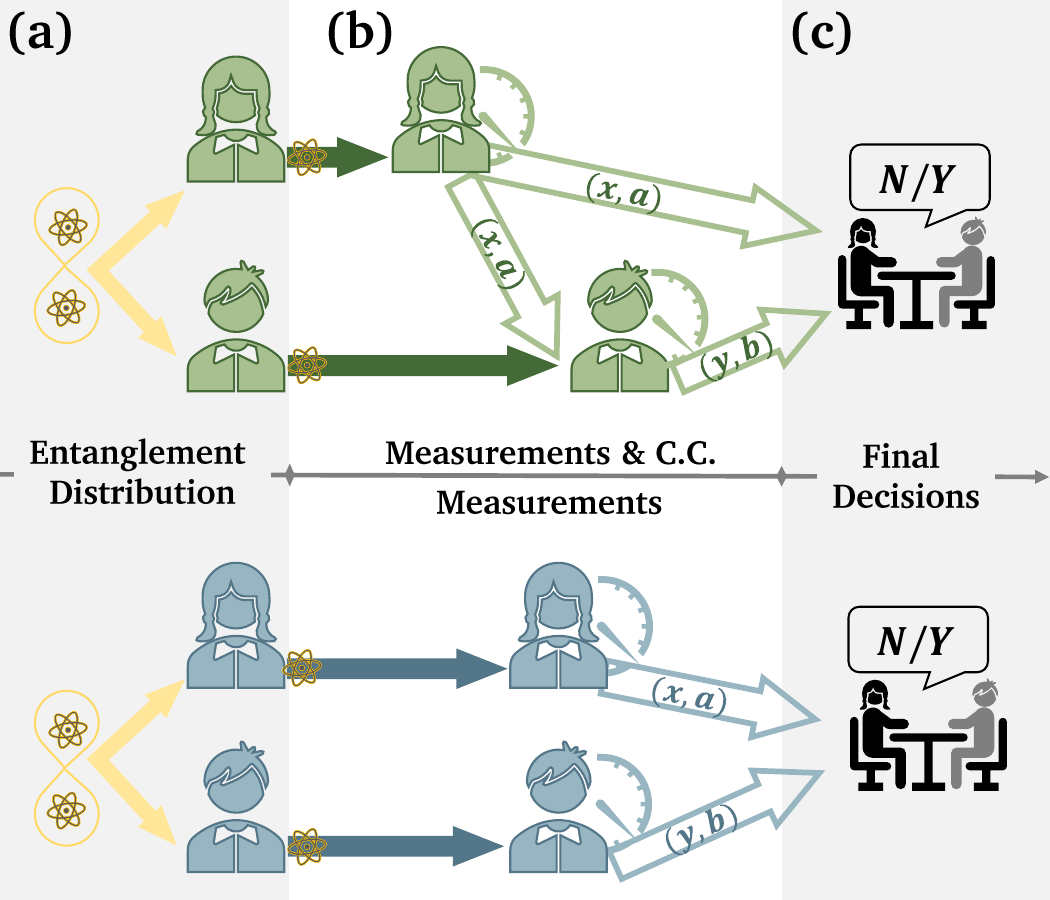}
    \caption{Outline of the 1-LOCC (top) and LO (bottom) protocols. (a) After state preparation, the photons are distributed to Alice and Bob by quantum channels. In Phase (b) measurements are conducted and classical information is communicated after that. In 1-LOCC, Alice chooses measurement setting $x$ randomly, obtains outcome $a$, and sends information $(x,a)$ through classical channels to Bob while keeping another copy for later use. Using this information, Bob then randomly chooses measurement setting $y$ and obtains outcome $b$. In LO, Alice and Bob choose $x$ and $y$ randomly and independently, obtaining outcomes $a$ and $b$ respectively. In phase (c) classical information $(a,b,x,y)$ from Alice and Bob is collected to determine whether the state is entangled (Y) or separable (N).}
    \label{fig:protocols}
\end{figure}

In our experiments, we employ two types of protocols in Fig.~\ref{fig:protocols}: LO and 1-LOCC protocols, where the communication goes from Alice to Bob.
The simplicity of the form $P(x,y,\gamma = N|a,b)$ masks the different theoretical and experimental constraints of these two protocols. The LO protocol stipulates that in each round of the experiment, Alice and Bob choose their measurement labels $x,y$ by sampling from a joint probability distribution $P(x,y)$, then record the outcomes $a,b$ to compute the final outcome $\gamma$. They effectively implement the POVM element
\begin{align}
	M_N^{LO} & =\sum_{a,b,x,y}P(N|x,y,a,b)P(x,y)(A_{x}^{a} \otimes B_{y}^{b}) \nonumber\\
	         & =\sum_{a,b,x,y}P_{LO}(x,y,N|a,b)(A_{x}^{a} \otimes B_{y}^{b}),\label{eq: M_lo}
\end{align}
with the other POVM element $M_Y^{LO}=\mathbb{I}-M_N^{LO}$.

The 1-LOCC protocol, on the other hand, is more demanding both theoretically and experimentally. At the beginning of each round, only Alice samples her measurement $x$ from a probability distribution $P(x)$. After she obtains the outcome $a$, she transmits both $x,a$ to Bob, who samples his measurement from the distribution $P(y|a, x)$. After Bob obtains his outcome $b$, they can compute the final outcome $\gamma$. The effective POVM is given by
\begin{align}
	&M_{N}^{1-LOCC}\nonumber\\
	 = &\sum_{a,b,x,y}P(N|x,y,a,b)(\mathbb{I}_A \otimes B_{y}^{b})P(y|a,x)(A_{x}^{a} \otimes \mathbb{I}_B)P(x)  \nonumber\\
	  =&\sum_{a,b,x,y}P_{1-LOCC}(x,y,N|a,b)(A_{x}^{a} \otimes B_{y}^{b}), \label{eq: M_locc}
\end{align}
with the other POVM element $M_Y^{1-LOCC}=\mathbb{I}-M_N^{1-LOCC}$.

The 1-LOCC protocol requires a steady supply of high-quality random numbers drawn from various complex distributions, the ability to store Bob's state while Alice carries out her operations and fast switching of Bob's measurements based on Alice's communication. Compared to the LO protocol, which only requires Alice and Bob to sample from a joint distribution $P(x,y)$, do these extra experimental costs bring an advantage in terms of having a lower probability of type-II error $p_2$?

\section*{From Conceptual Setup to Experimental Constraints and Back}
To answer the question above, we need to allow the LO and 1-LOCC versions of Eq.~\eqref{eq:sdp} to share as much experimental resources as possible, while comparing the two resulting $p_2^*$. In Eq.~\eqref{eq:sdp} the experimental constraints are inputs to the problem: the target state $\rho$ to be certified and the allowed projective measurements $A^a_x$ and $B^b_y$. Sharing these resources keeps the state and the allowed measurements the same for both protocols, while the probabilistic instructions $P(x,y,\gamma=N|a,b)^*$ may differ.

In an earlier work~\cite{QuanPereGame21}, some of us showed that there are states where there is an advantage of using 1-LOCC over LO. However, for many 2-dimensional and simple, symmetric 3-dimensional states, the type-II error probability for the two protocols is so close that implementing the classical communication does not bring any meaningful advantage. Therefore, to design our experiment, we need to find a state and a set of measurements which show a meaningful difference between the two $p_2^*$, under constraints imposed by modern quantum optic experiments.

\begin{figure*}[tbph!]
    \centering
    \includegraphics[width=1\textwidth]{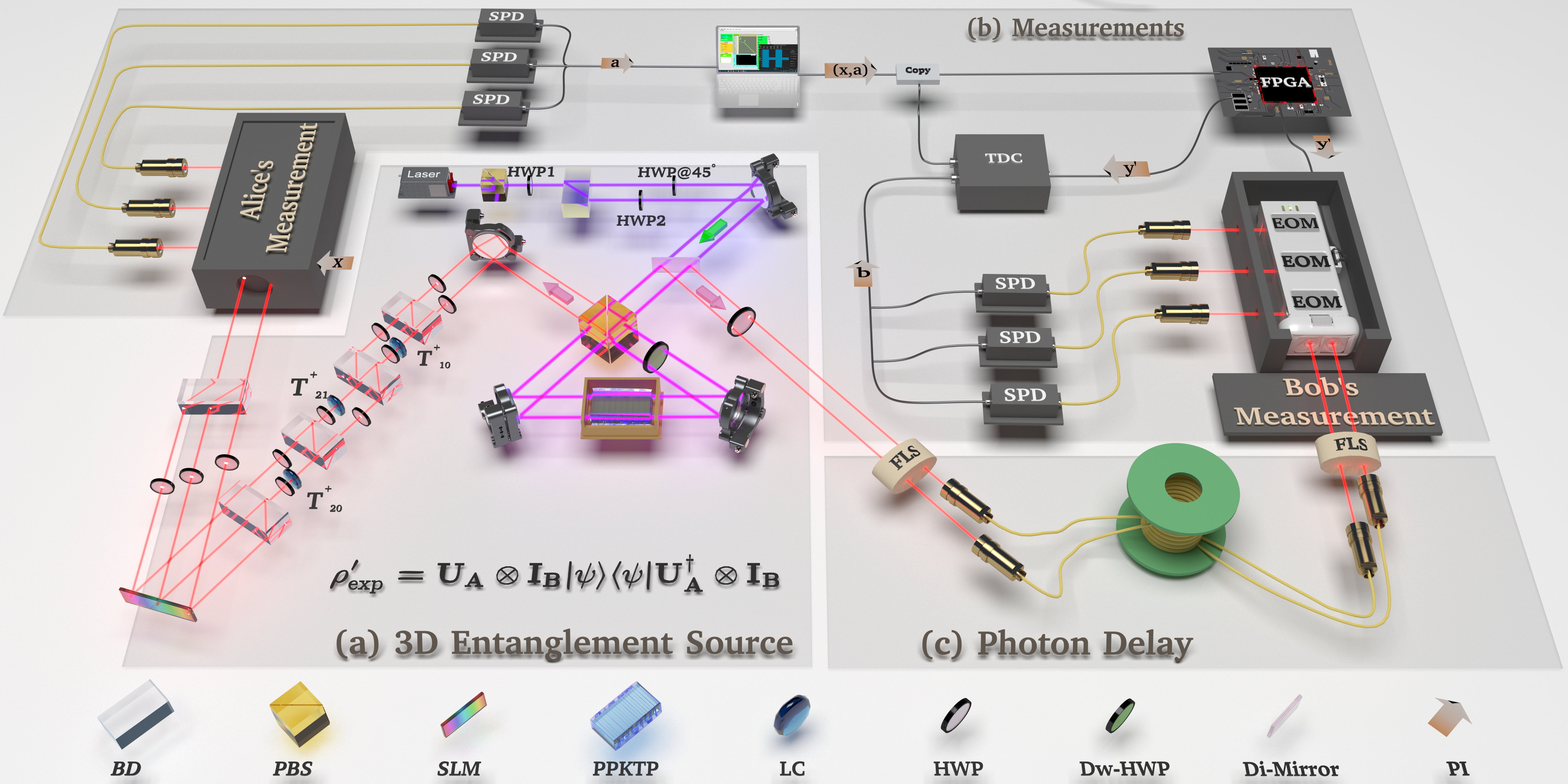}
    \caption{Experimental setup. \textbf{(a) 3D Entanglement Source: }3D entangled photons are generated via spontaneous parametric down-conversion (SPDC) in a PPKTP crystal, pumped by a 404~nm continuous-wave laser with 20~mW power, using polarization-path hybrid encoding~\cite{DenseCode18,OptimDectLOCC21,HuXing20}. By adjusting HWP1-2, $|\psi\rangle$ in Eq.~\eqref{setup_state} is prepared, which then undergoes Alice's unitary operation $U_A$ to produce the experimental target state $\rho'_{\text{exp}}$. The $ U_A $ is further decomposed into three two-level unitary transformations $T^{\dagger}_{ij} $, with details provided in the Supplementary Information. \textbf{(B) Measurements: }This part is responsible for entangled photons detection and the implementation of the LOCC protocol. Both Alice’s and Bob’s measurement setups allow switching among three measurement bases. Bob’s measurement basis $y'$ is switched via three FPGA-controlled EOMs, where $ y' $ is influenced by Alice’s measurement basis $ x $ and results $ a $. In contrast, $ x $ is switched by the experimenter, with $ x $ determined through random number sampling. The computer and FPGA handle classical communication (acceptance/rejection), store random number tables, and control the EOM. The TDC is responsible for data acquisition and analytical decision-making (Y or N). \textbf{(c) Photon Delay: }Two 50~m fibers are used for photon delay to ensure that Bob's measurement is prepared before the arrival of the photon. BD: beam-displacer; PBS: Polarizing Beam Splitter; SLM: Spatial Light Modulator; PPKTP: Periodically Poled KTP; LC: Liquid Crystal; HWP: Half Wave Plate; Dw-HWP: Dual-wavelength Half Wave Plate; Di-Mirror: Dichroic Mirrors; PI: Probabilistic Instructions. FLS: Fiber Locking System;  FPGA: Field Programmable Gate Array Chip; SPD: Single Photon Detector; TDC: Time-to-Digital Converter. Details of the experimental setup, Measurements, and FLS are provided in the Supplementary Information.}
    \label{fig:setup}
\end{figure*}

Drawing on our experience with high-dimensional quantum state engineering~\cite{DenseCode18,HuXing20,Hu2021Highnoise}, we implement a polarization-path hybrid entanglement source depicted in Fig.~\ref{fig:setup}, which allows us to prepare arbitrary qutrit-qutrit states of the form
\begin{align}
\rho'_{\rm exp}=U_A\otimes I_B|\psi\rangle\langle\psi|U_A^{\dagger}\otimes I_B,\label{setup_state}
\end{align}
where $|\psi\rangle$ is provided in the Methods. The preparation of the quantum state described above requires precise angular adjustments of optical components in Fig.~\ref{fig:setup}. For complex three-dimensional quantum states, determining these angular configurations is non-trivial. To achieve high-precision preparation of quantum states that demonstrate LOCC advantages, we use a systematic method that analytically maps target quantum states to angular parameters by solving a set of equations. This approach simplifies state preparation to configuring optical devices based on the computed angular parameters. Technical details are provided in the Methods. Additionally, the measurements are assumed to be projections on the computational basis, the Fourier basis, or the Gell-Mann basis, which can all be efficiently implemented using our setup. 

After taking these experimental constraints into account, our initial problem Eq.~\eqref{eq:sdp} is transformed into a variational optimization problem with a convex optimization subproblem for the variables $\{e_1, \mathcal{X}\}$:
\begin{equation}
    \begin{aligned}
        \max_{\{e_1, \mathcal{X}\}} \quad  & G(\rho,p_1) = p_2^{LO*}(\rho,p_1)-p_2^{LOCC*}(\rho,p_1)
    \end{aligned}, \label{eq:max_gap}
\end{equation}
where $\mathcal{X} = \{\phi,\theta \in \mathbb{R},\ \bm{\lambda}_A \in \mathbb{R}^{3^2}\}$ is the parameter space of state preparation~(\ref{eq:max_state}) and $e_1$ is the parameter of false-positive error probability, which is formulated as $p_1=({\rm tanh}(e_1)+1)/2$.

While $ G(\rho,p_1)$ in Eq.~\eqref{eq:max_gap} quantifies the advantage of one-way communication in entanglement detection for the state $\rho$ with given type-I error probability $p_1$, finding optimal or near-optimal solutions to Eq.~\eqref{eq:max_gap} poses a challenge. Using na\"{i}ve gradient descent algorithms requires evaluating dozens of convex optimization problems for each gradient evaluation. The algorithms also get trapped easily in local minima. To overcome these obstacles, we devised an optimization algorithm based on deep generative neural networks to solve this problem. The resulting Variational Generative Optimization Network (VGON) effectively outperforms na\"{i}ve gradient descent algorithms by greatly reducing the optimization time for Eq.~\eqref{eq:max_gap}~\cite{VGON24}.

After a near-optimal solution to Eq.~\eqref{eq:max_gap} has been found by VGON, it gives us the target state $\rho^{*}$ (The expression and the angles required for preparation are given in the Supplementary Information), the corresponding $p_2^*$ values as $p_2^{LO*}=0.0944$ and $p_2^{LOCC*}=0.0283$, with $p_1=0.7481$, resulting in a gap of $G=0.0661$, and two sets of instructions $P_{1-LOCC}^*(x,y,N|a,b), \; P_{LO}^*(x,y,N|a,b)$. The explicit state, corresponding experimental parameters, and resulting instructions are detailed in the Supplementary Information. Each instruction set is made of smaller distributions: for the 1-LOCC set, we have $P^*(x),\;P^*(y|a,x),\;P_{1-LOCC}^*(N|x,y,a,b)$; for the LO set, we have $P^*(x,y)$ and $P_{LO}^*(N|x,y,a,b)$. In our experiment, samples from these probability distributions are stored in an FPGA chip, which serves as its nerve center.

\section*{Experimental Setup and Result Analysis}
We closely follow our entanglement detection scenario when implementing the 1-LOCC protocol. At the beginning of each experimental round, the entangled state of the target $\rho^*$ is prepared by setting the HWPs and LCs in Fig.~\ref{fig:setup}. Alice then measures $x \in \{0, 1, 2\}$ and obtains outcome $a$. She communicates $x,a$ to Bob, who samples $y$ from $P(y|a,x)$ and commands the FPGA to configure the EOMs to prepare the measurement $y$. Meanwhile, his photon is temporarily stored in a simple quantum memory until his measurement is ready. The EOMs have a repetition rate of $10~\text{MHz}$ and the quantum memory takes the form of a photon delay mechanism, implemented using two $50~m$ fibers. After Bob obtains his outcome $b$, the Time-to-Digital Converter (TDC) records and analyzes the coincidence count $(a, b)$. The final outcome (coincidence count) is randomly rejected with probability $ P^*_{1-LOCC}(N|a, b, x, y) $, completing one round of the 1-LOCC process. Repeating for a few million rounds allows us to collect statistics of coincidence counts to estimate the false negative error probability $ p_2^{1-\text{LOCC}*} $ for the 1-LOCC protocol.
Experimental details can be found in Supplementary Information.

The stability of the entanglement source and our experimental goal stipulates that the best way to control for the effect of classical communication is to keep as much of the setup as possible and shuffle the rounds of LO and 1-LOCC, so that unstable experimental environments and devices will affect the two protocols equally. Therefore, instead of sampling $x,y$ from $P^*(x,y)$, we replace $P(y|a,x) $ with $P(y|x) $, from which we get $P(x,y) = P(y|x)P(x)$. The distributions $P(x)$ for both protocols are combined into a single distribution $P_{m}(x)$. In each experimental round, the settings $x$ are sampled from $P_{m}(x)$ and used in both protocols. To allow TDC to track which protocol is implemented in each round, $P(y|x)$ and $P(y|a,x)$ are merged by encoding the settings $y$ of LO as $y'\in\{0,1,2\}$ and those from 1-LOCC as $y'\in\{3,4,5\}$. For each $x$, the FPGA switches Bob’s measurement by sampling from the merged distribution $P_{m}(y'|a,x)$, given $a$ and $x$ as inputs. For $y'\in\{0,1,2\}$, i.e. the LO protocol is implemented, $P_{m}(y'|a,x)$ is the same for all $a$, ensuring Alice's measurement outcome does not influence Bob's choice of measurement settings. A single experimental round is complete after the TDC records a coincidence event $(a,b,x,y')$. The distributions mentioned above are given in Supplementary Information.

\begin{figure}[tbph!]
    \centering
    \includegraphics[width=0.4\textwidth]{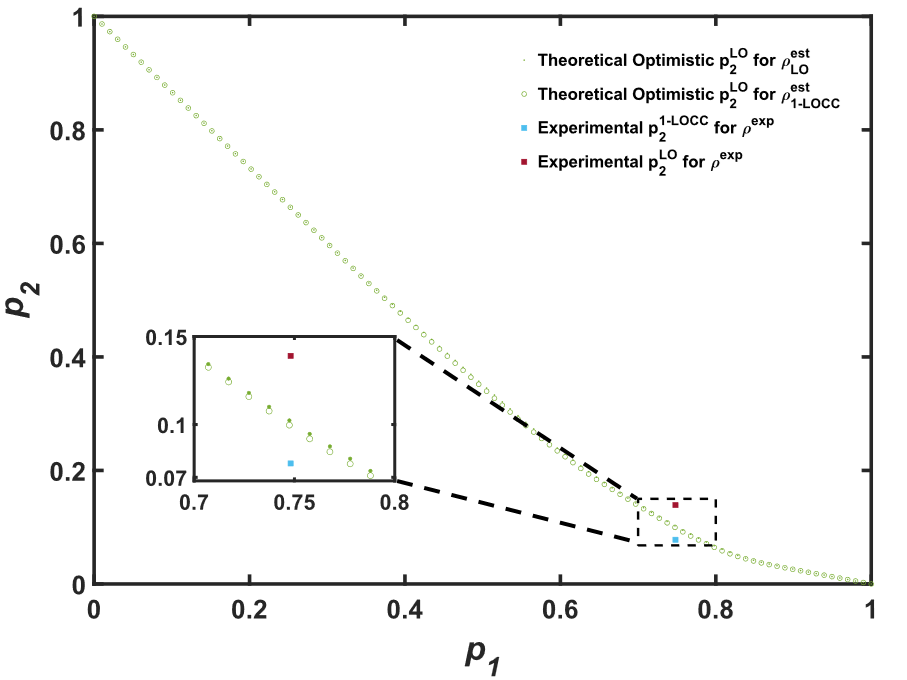}
    \caption{Comparison of optimistic false-negative probabilities in LO with experimental and theoretical values under 1-LOCC. Green dots and circles represent optimistic $p_2$ values for the estimated states $\rho^{est}_{LO}$ and $\rho^{est}_{1-LOCC}$ under LO scenario. The experimental $p_2$ value (blue square) in 1-LOCC scenario is below the optimistic estimate LO curves. The red square represents the experimental $p_2$ value in LO scenario.}
    \label{fig:inner}
\end{figure}

The source in Fig.~\ref{fig:setup} generates 1200 pairs of entangled photons per second, with a coincidence efficiency of $~10~\%$ in an 8 ns coincidence window. Total data collection time is 28 minutes, resulting in about 1 million experimental rounds per protocol. After data collection and analysis, the experimental false-negative error probabilities of 1-LOCC and LO are 0.0779 and 0.1390, resulting the gap in Eq.~\eqref{eq:max_gap} of $0.0611$, indicating the advantage of classical communication in reducing statistical errors. However, to confirm the advantage of classical communication over all LO protocols for the states generated by our setup, the experimental $p_2^{1-LOCC}$ value should be smaller than $p_2^{LO}$ even under the most optimistic estimation. This, in turn, requires us to estimate the states $\rho_{est}^{LO}$ and $\rho_{est}^{1-LOCC}$ from raw experimental data. The measurement bases used in our protocols are far from tomographically complete, therefore we use an SDP to find the states which deviate the least from raw experimental data under the $l^2$ norm. With $\rho_{est}^{LO}$ and $\rho_{est}^{1-LOCC}$, we obtain optimistic estimations of $p_2^{LO}$ using a set of random product states and fixed values for $p_1$. Fig.~\ref{fig:inner} shows that even under these optimistic estimations, the experimental $p_2$ value for LOCC (blue square) still outperforms that of LO (green curve). The Methods section contains additional details on how $\rho_{est}^{LO}$, $\rho_{est}^{1-LOCC}$ and the green curve are obtained.

 This work experimentally demonstrates the advantage of extra classical information in reducing the statistical error in entanglement certification protocols. By integrating protocols with and without classical communication in the same experimental setup, the advantage of classical information was fairly identified and quantified using short-lived memory techniques and high-speed measurement switching devices.

The experimental results align with theoretical predictions, validating the robustness of the 1-LOCC advantage even in the presence of noise. This proof-of-principle experiment highlights the potential of 1-LOCC protocols as scalable and efficient tools for detecting quantum entanglement, offering significant advancements for quantum networks and the quantum internet. Future work should focus on extending these techniques to multipartite and higher-dimensional systems, refining noise models, and exploring adaptive measurement strategies to enhance their practical applications in quantum technologies further.

\section*{Acknowledgments}
This work was supported by the NSFC (No.~12374338, No.~11904357, No.~12174367, No.~12204458, No.~62250073, No.~62272259), the Sichuan Provincial Key R\&D Program (2024YFHZ0371), the National Key R\&D Program of China (2021YFE0113100), the Innovation Program for Quantum Science and Technology (No. 2021ZD0301200),  Anhui Provincial Natural Science Foundation (No. 2408085JX002), Anhui Province Science and Technology Innovation Project (No. 202423r06050004), China Postdoctoral Science Foundation (2021M700138).

\bibliography{AdvantageCC}
\clearpage

\appendix
\section{Additional Experimental Details}\label{sec:methods}
The quantum state $ |\psi\rangle $ in Eq.~\ref{setup_state} has the following form:  

\begin{align}
    |\psi \rangle = \sin \frac{\theta}{2} \cos \frac{\phi}{4} |00 \rangle +
    \sin \frac{\theta}{2} \sin \frac{\phi}{4} |11 \rangle+
    \cos \frac{\theta}{2}|22 \rangle,\label{eq:max_state}
\end{align}
where $ \phi \in [0,2\pi) $ and $ \theta \in [0,\pi] $. This state is derived from the simultaneous solution of the pump beam state and the Jones calculus for HWP1 and HWP2 in Fig.~\ref{fig:setup}, and is generated via SPDC.

The local unitary on Alice's side $U_A$ is parametrized by a set of $3^2$ linearly-independent skew-Hermitian matrices $\{T_j\}$~\cite{Reck1994}:
\begin{align}
    U_A = \exp(\sum_{j = 1}^{3^2}\lambda_j T_j),\label{eq:UA}
\end{align}
where $\lambda_j$'s are $3^2$ parameters, denoted as $\bm{\lambda}_A$. 
The \(\{T_j\}\) and \(\bm{\lambda}_A\) can be expressed in terms of the operational matrices of HWP, LC, and SLM in \( U_A \), followed by an analytical solution.  

By jointly solving Eqs.~\eqref{eq:max_state}-\eqref{eq:UA}, we establish a mapping between the quantum state in Eq.~\ref{setup_state} and the angular configurations of optical components. The angles required for this experiment are provided in the Supplementary Information.

Event-ready measurements for the two protocols require sampling random numbers from three distributions: $ p_m(x) $, $ p_m(y|a,x) $, and $ p^*(N|a,x,b,y) $, after the slight tweaking of one of the protocols. In our experiments, all random numbers were pre-generated using Mathematica. A total of 172 random number tables were generated: one for $ x \in \{0,1,2\} $ (RT$x$), nine for $ y' \in \{0,1,2,3,4,5\} $ (RT$y'$), and 162 for $ N \in \{0,1\} $ (RT$N$). The explicit distributions can be found in the Supplementary Information. Before measurement, the runtime for each $ x $ setting is determined by the digit counts in RT$x$ (e.g., if RT$x$ contains 10 zeros, $ x=0 $ is measured for 10 seconds). RT$y'$s are stored in the FPGA, while RT$N$s are stored in the TDC.  

The measurement begins with the experimenter configuring the setups and announcing the choice of $ x $, i.e. $x$ is known for all devices. Alice’s result $a$ for each detected entangled photon is transmitted to the FPGA and copied to the TDC (e.g., Fig.~\ref{fig:setup}). Using each $(x,a)$, the FPGA selects $ y' $ from RT$y'$ and configures Bob’s measurement basis by EOMs as $ y = y' \mod 3 $. Bob's result ($ b  \in \{0,1,2\}$) corresponds to B1-B3, and sends $ b $ to the TDC. The measurement configuration settings are detailed in the Supplementary Information.

At this stage, the TDC knows $(a,b,x)$ but lacks $ y $, preventing it from selecting RT$N$ to sample $ N $. To resolve this, five additional FPGA outputs are connected to the TDC: three for $ y $ and two indicating whether the measurement corresponds to the 1-LOCC or LO protocol. Using $(\text{LO}/\text{1-LOCC},a,b,x,y)$, the TDC selects a random number from RT$N$ to decide whether to reject the measurement. The experimenter counts four-body coincidence counts $[\text{LO}/\text{1-LOCC},a,b,y]$ using the TDC. The $ x $ informs the TDC of the measurement timing but is not involved in coincidence counts. The experiment is complete after setting $ x $ three times and obtaining all measurements.  

This process ensures accurate, simultaneous measurement of both protocols under identical noise conditions.

\section{Estimating the experimental states and optimistic $p_2$ for LO}
After data collection and analysis, we notice that the experimental false-negative values increased, while the gap between them remained quite large, likely due to state deviations arising from experimental imperfections. These deviations make it unreasonable to assume that theoretical expectation values perfectly align with the raw experimental data. Therefore, we estimate $\rho_{exp}$ by finding the best state $\rho_{est}$ which minimizes deviations from raw data quantified by the sum of squared error between nonzero raw expectation values $E^{exp}$ and corresponding theoretical estimations. Here $E^{exp}=\{E^{exp}_{(a,b,x,y)}\}_{(a,b,x,y)\in I'}$, and $I' = \{(a,b,x,y)|P(x,y,N|a,b)^*\neq 0\}$. The resulting optimization problem is given by 

\begin{align}
    \min_{\rho_{est}} \quad &
    \| E^{exp}_{(a,b,x,y)}-E_{(a,b,x,y)} \|^2 \quad {(a,b,x,y) \in I'} \nonumber\\
    \text{s.t. } & E_{(a,b,x,y)}=tr(\rho_{est} (A_{x}^{a}\otimes B_{y}^{b})) \quad {(a,b,x,y) \in I'} \nonumber\\
                & tr(\rho_{est})=1; \quad \rho_{est} \geq 0
    \label{eq:est_state_sdp}
\end{align}
where $\{E_{(a,b,x,y)}\}_{(a,b,x,y) \in I'}$ represents the theoretical expectation values for the estimated state $\rho_{est}$, corresponding to the raw data of local measurement pairs in the $E^{exp}$.  The values of the objective functions obtained by cvx using the solver Mosek~\cite{mosek}, quantifying the minimum amount of correction needed to make the raw experimental data consistent with physical density matrices, are both on the order of $10^{-4}$ for $E^{exp}_{LO}$ and $E^{exp}_{1-LOCC}$.

Taking the value of $p_2$ as a metric, the estimated states $\rho_{est}^{LO}$ and $\rho_{est}^{1-LOCC}$ more accurately represent $\rho_{exp}$ than $\rho^*$, while the gaps are also consistent with the experiment (see Supplementary Information). To confirm the advantage of classical communication, we obtain an optimistic estimation of $p_2$ for LO protocols based on the estimated states.

Our conceptual setup, Eq.~(\ref{eq:sdp}), corresponds to an outer approximation of the set of separable states via the dual of the first level of the DPS hierarchy. It can be seen as a pessimistic estimation of $p_2$. On the other hand, an optimistic estimation can be obtained by solving the following linear programming problem, which we solve using Gurobi~\cite{GurobiOptimization2024}:
\begin{equation}
	\begin{aligned}
		\min_{P(x,y,\gamma|a,b)} \quad & p_2^{LO}  \\
		\text{subject to }\quad       & {\rm tr}(M_N\rho_{est})=p_2,     \\
		                                 & tr(M_Y\rho_i)\leq p_1, i=1,...,d,   \\
		                               & M_{N}=\sum_{a,b,x,y}P(x,y,\gamma=N|a,b)(A_{x}^{a} \otimes B_{y}^{b}) \\
					                   &M_{Y}=\mathbb{I}-M_{N}\\
		                               & \sum_{\gamma}P(x,y,\gamma|a,b)=P(x,y),\quad \sum_{x,y}P(x,y)=1.
	\end{aligned}\label{eq:lp_inner}
\end{equation}
$\{\rho_i\}_{i=1}^{d},d=100,000$ are random product states that approximate the separable set from the inside. We plot the optimum solutions to Eq.~(\ref{eq:lp_inner}) against 100 different values of $0\leq p_1 \leq1$, which gives the green curve in Fig.~(\ref{fig:inner})

\section{Optimized state, protocols, and probabilistic instructions}
Our research aims to experimentally demonstrate the advantages of LOCC in entanglement detection. A more efficient solution to Eq.~\eqref{eq:max_gap} will help identify quantum states that offer greater advantages—those that are easier to measure and can be experimentally prepared. To achieve this, we have designed a variational generative optimization network (VGON)~\cite{VGON24}. The resulting state $\rho^{*} = \ket{\phi^*}\bra{\phi^*}$, generated by the VGON, is presented below. Although it is neither regular nor symmetric, a strategy for its experimental preparation can be developed by solving a system of equations. The experiment strategy are shown in Table~\ref{Tab:angles}. 
\begin{equation}
    \ket{\phi^*} = 
\begin{pmatrix}
    -&0.2707 - 0.2859i,\\
     &0.2435 + 0.2818i,\\
    -&0.1624 + 0.1675i,\\
     &0.0932 + 0.1845i,\\
    -&0.1873 - 0.1474i,\\
    -&0.2944 + 0.2817i,\\
     &0.3284 - 0.2914i,\\
     &0.3043 - 0.3131i,\\
    -&0.0182 + 0.0503i
\end{pmatrix} \label{eq:state}
\end{equation}

The experimental preparation strategy for 
$\rho^{*}$ is shown in Table~\ref{Tab:angles} for various waveplate angles. The corresponding waveplates in the table are linked to those depicted in Fig.~\ref{fig:setup}.

The $\rho^{*}$'s minimal false-negative error probabilities $p_2^{LO*}$, $p_2^{1-LOCC*}$ and advantage $G$ are presented in Table~\ref{Tab: data}. Additionally, the optimal instructions for the LO and 1-LOCC scenarios are detailed in Table \ref{Tab: Pxyn_ab}. The distributions associated with the construction of measurement protocols of LO and 1-LOCC $M_{N}^{*LO}$ and $M_{N}^{*1-LOCC}$ are in Table~\ref{Tab: decision} and Table~\ref{Tab: PN_abxy}. These probabilities all come from solving Eqs.~\eqref{eq:sdp}-\eqref{eq: M_locc}.

\section{Experimental Measurement Settings}\label{exp_measurement}
During the measurement stage, for simplicity, the sets of local measurements available to Alice and Bob are kept the identical and fixed, in the configuration of the experimental routine computational basis shown in Eq.~\eqref{eq: basis}, the Fourier basis, and the Gell-Mann basis, which are also used as known information to solve for $M^{*}_{\text{LO}_N}$ and $M^{*}_{1-\text{LOCC}_N}$.

\begin{figure*}[htbp!]
\centering
\includegraphics[width=1\textwidth]{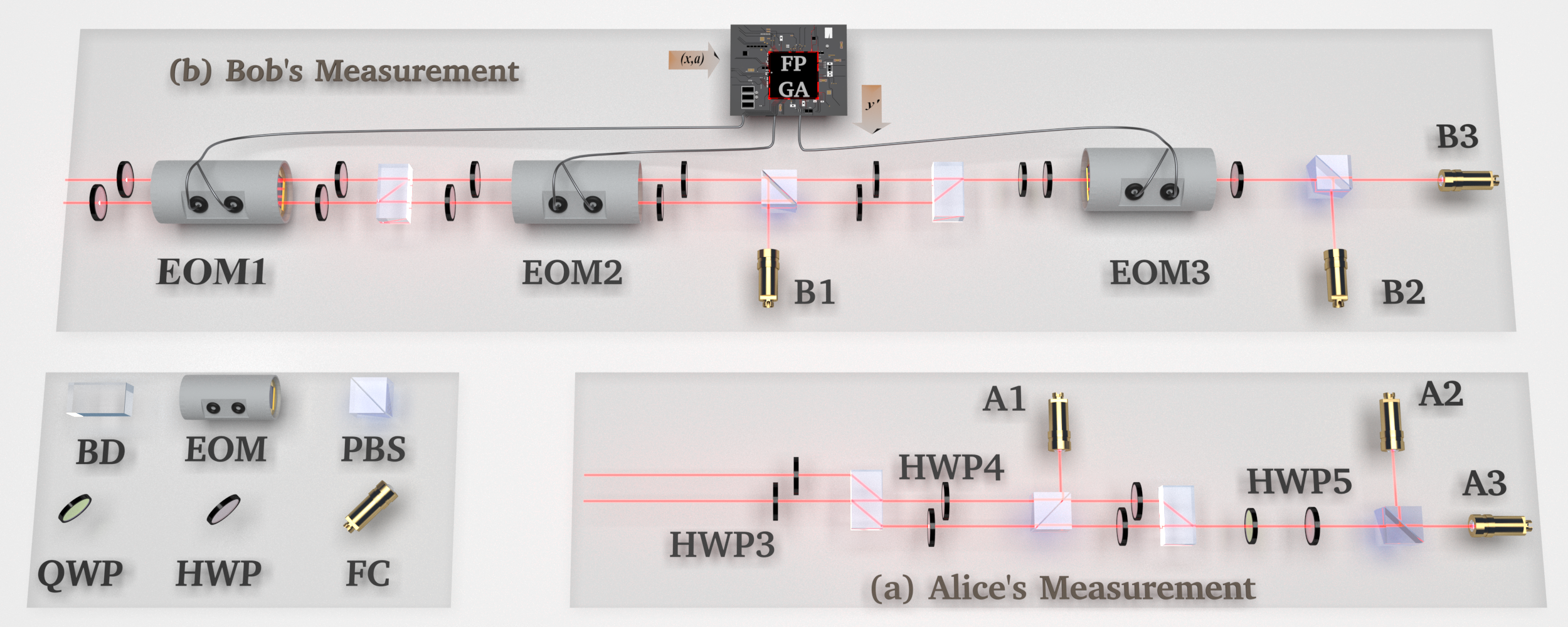}
\caption{\textbf{Experimental setup of local measurements.} Switching between the three measurement settings can be accomplished for both Alice and Bob measurements. \textbf{(a) Alice's Measurement}, the implemented bases are the computational basis, Fourier basis, and Gell-Mann basis, which are realized by tuning HWP3, HWP4, and HWP5 to the angles in table ~\ref{Tab: angle_measure}. \textbf{(b) Bob's Measurement}, real-time switching is realized by the fast configuration switching of extra EOM1-EOM3, with a repetition frequency of $ 10~\text{MHz} $. Specifically, the two-level modes of EOM (offset or pulse) are switched by different Volts. The correspondence between the bases and configuration of EOMs is shown in the table.~\ref{Tab: voltage}.}
\label{fig: Measurement}
\end{figure*}

Alice’s measurements are implemented using HWP3-HWP5, as shown in Figure \ref{fig: Measurement}(a), with the corresponding angles specified in Table~\ref{Tab: angle_measure}. The dynamic nature of the 1-LOCC requires Bob to first await information from Alice and then rapidly change basis in response to her signals, which necessitates the short-lived memory and real-time response on Bob's side. The real-time switching among the three measurement bases is initiated by Alice's signals and facilitated by various configurations of extra electro-optical modulators (EOMs) controlled by an FPGA. The experimental setup is depicted in Fig.~\ref{fig: Measurement}(b), with the configurations of EOMs details provided in Table.~\ref{Tab: voltage}. The memory duration is determined by the response times of the EOMs and FPGA, which are 120 ns and 50 ns, respectively. To achieve the necessary delay, we employed two 50-meter optical fibers, providing a maximum delay time of approximately 245 ns, ensuring that Bob's measurement basis was adequately prepared before the arrival of the photon to be measured.

\begin{equation}
    \begin{aligned}
        A_1 =&\{|0\rangle,\ |1\rangle,\ |2\rangle\},\\
        A_2 = &\{\frac{1}{\sqrt{3}}(|0\rangle+e^{-i \frac{2 \pi}{3}}|1\rangle+e^{-i \frac{-2 \pi}{3}}|2 \rangle), \\
        &\frac{1}{\sqrt{3}}(|0\rangle+e^{-i \frac{-2 \pi}{3}}|1\rangle+e^{-i \frac{2 \pi}{3}}|2 \rangle) , \\
        &\frac{1}{\sqrt{3}}(|0\rangle+|1\rangle+|2\rangle)\}, \\
        A_3 =&\{\frac{1}{\sqrt{2}}(|1\rangle-|2\rangle),|0\rangle,\frac{1}{\sqrt{2}}(|1\rangle+|2\rangle)\}.\\
    \end{aligned}\label{eq: basis}
\end{equation}

\section{The Photon Delay Mechanism}\label{sec:FLS}
During 1-LOCC, it is critical that Bob’s photon arrives after the FPGA’s electrical output to ensure that his measurement basis is ready. In our actual experiment, the response time of FPGA and EOM is approximately $170~ns$, so we introduce a $50~m$ fiber delay, which provides a waiting time of $250~ns$ for the optical signal. But this will bring about a problem, i.e., the phase between two paths is not stable without active feedback. So we use an 830nm light as reference light to lock the phase of entanglement photon which is 808nm in the fiber locking system~\cite{HuXing20}. 

\begin{figure}[tbph]
    \centering
    \includegraphics[width=0.45\textwidth]{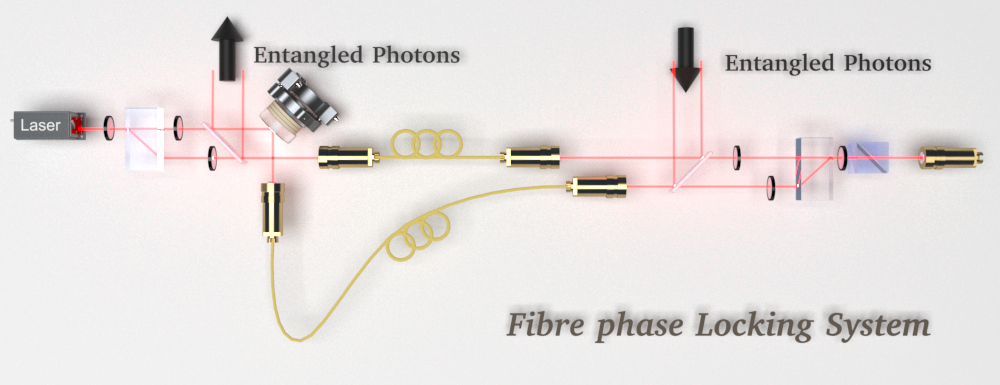}
    \caption{Fiber Locking System. The reference light is divided into two beams by one beam displacer and recombined by another beam displacer, composing a Mach-Zehnder interferometer with two $50m$ optical fibers. A mirror of a piezoelectric ceramic (PZT) is used to change length of the optical path to lock the phase of two paths after $50m$ optical fiber according to the signal of reference light.}
    \label{fig: FLS}
\end{figure}

As shown in Fig~\ref{fig: FLS}, two narrowband filters are used to separate entangled photons and reference light.  The reference light is emitted by a semiconductor laser and then passes through a beam displacer, divided into two beams. To reduce disturbance to entanglement photons, the reference light is attenuated to $10^{6}$ Hz and finally detected by a single photon detector. The reference light in upper path is reflected by the mirror of a piezoelectric ceramic material (PZT) to change length of the optical path. After that, reference light in two paths is coupled into one end of 50m optical fiber and then emitted from the other end. Two beam displacers form a Mach-Zehnder interferometer. We use the PZT to adjust the position of the mirror according to the signal of the Mach-Zehnder interferometer to lock the phase of the optical fiber.

\begin{figure}[tbph]
	\centering
	\includegraphics [width=0.48\textwidth]{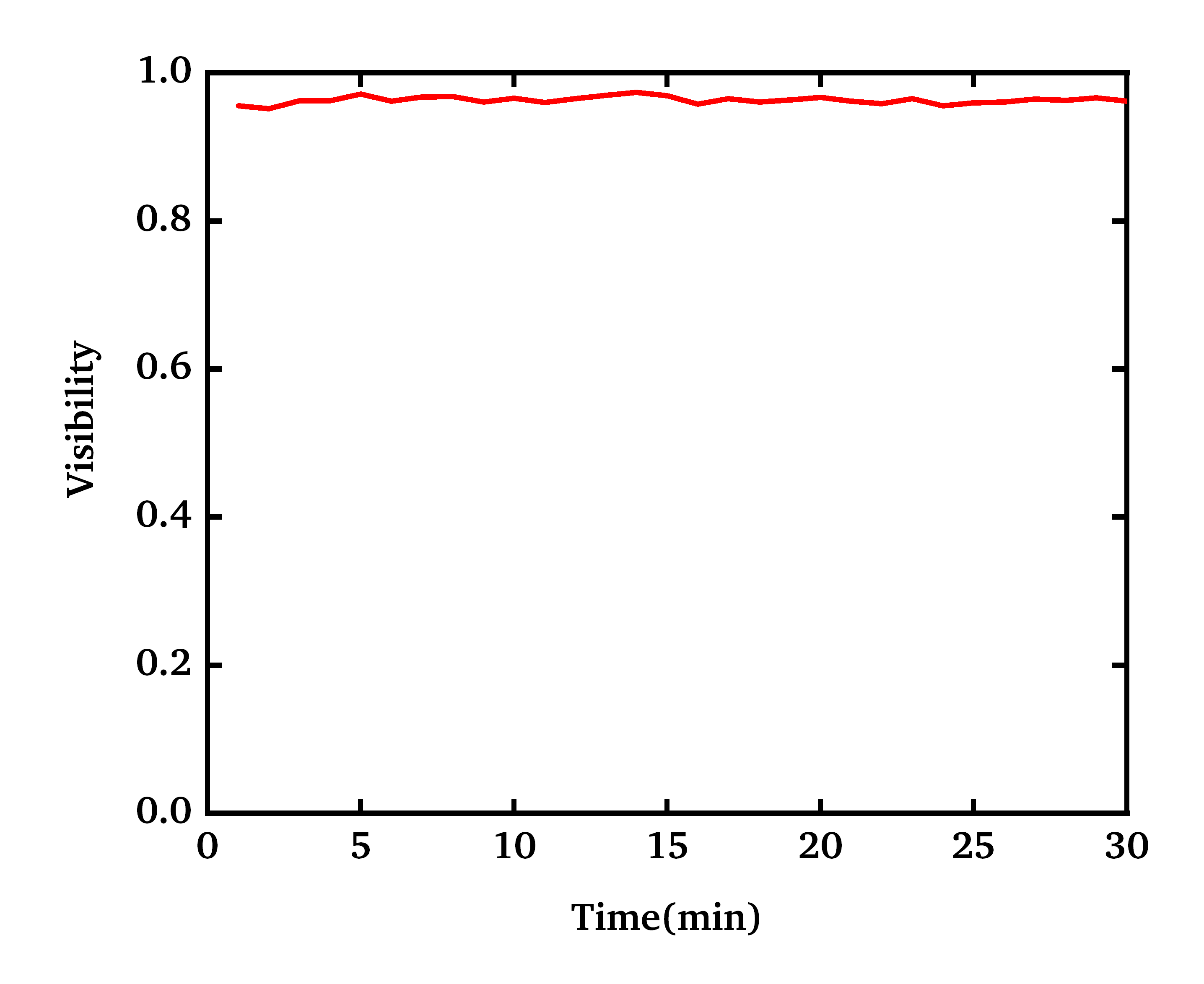}
	\caption{ Interference visibility between two paths with active feedback. The average entanglement visibility of the two path interference in half an hour is $V = 0.963$.}
	\label{fig:visibility}
\end{figure}

As shown in Fig.~\ref{fig:visibility}, we measure the visibility of entanglement photons between two paths to test the feedback system.

\section{Additional Analysis of Experimental Results}\label{sec:results}
After completing the experiments, Tab.~\ref{Tab: data} shows that the advantage of extra classical information has been experimentally demonstrated, as the minimal false-negative probability for 1-LOCC is smaller than that of LO for the same false-positive probability of 0.7481. While the experimental $p_2$ values deviate from theoretical predictions, the discrepancy in the advantage is much smaller, indicating that errors in $p_2$ have minimal impact on the observed advantage. Using the expectations of optimized protocols $M^*_{N}$ as a metric, Tab.\ref{Tab: data} shows that the false-negative errors for the estimated states $\rho_{est}^{LO}$ and $\rho_{est}^{1-LOCC}$ in both LO and 1-LOCC scenarios are closer to experimental false-negative errors than theoretical ones, while maintaining a low advantage discrepancy. Additionally, the fidelity between $\rho_{est}^{LO}$ and $\rho_{est}^{1-LOCC}$ is $94.34\%$, indicating that the shuffle strategy stabilizes the state for both LO and 1-LOCC.

\begin{figure}[tbph!]
    \centering
    \includegraphics[width=0.46\textwidth]{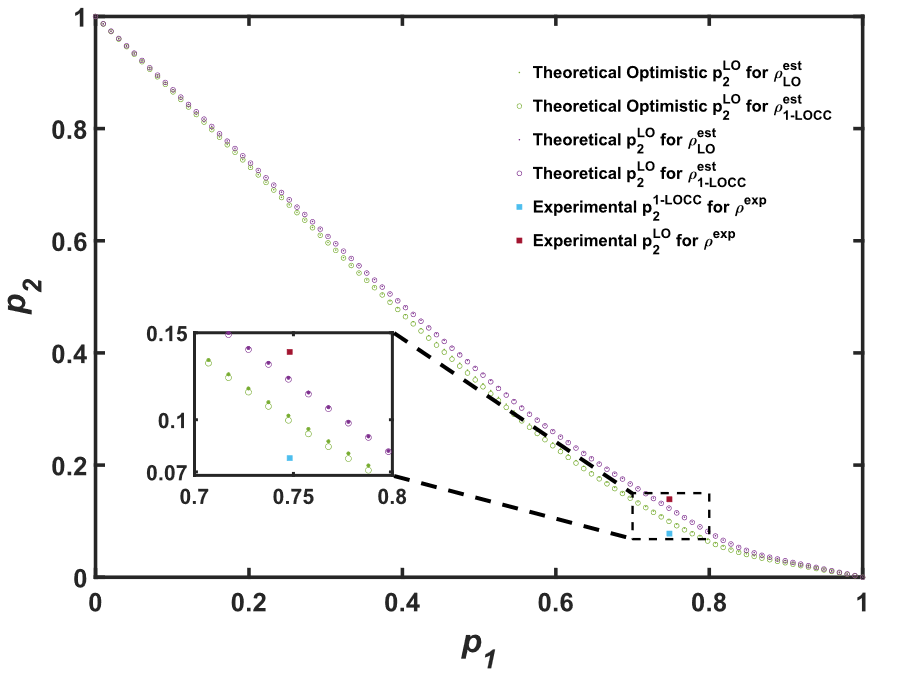}
    \caption{The comparison of false-negative error probabilities $p_2$s for false-positive error probabilities $p_1 \in [0,1]$. The purple dots and circles represent approximations of $p_2$ in Eq.~(\ref{eq:sdp_outer})for estimated states $\rho_{est}^{LO}$ and $\rho_{est}^{1-LOCC}$ under the LO scenario. The green markers show optimistic approximations in Eq.~(\ref{eq:lp_inner}) for the estimated states. Both approximations for the two estimated states almost overlap, demonstrating that the shuffle strategy stabilizes the state for both LO and 1-LOCC. The blue square indicates experimental $p_2$s values under the 1-LOCC scenario (circle). It can be found that the experimental 1-LOCC $p_2$ (blue square) is below both optimistic estimates and pessimistic ones of LO. The red square represents the experimental $p_2$ value in the LO scenario.}
    \label{fig:inner_outer}
\end{figure}

The advantage of 1-LOCC is evident from the smaller $  p_2^{\text{1-LOCC}}$ compared to $ p_2^{\text{LO}} $ at the same $p_1$. Experimental results confirm this with a 0.0611 gap in Eq.~\eqref{eq:max_gap}, despite some deviations from theoretical predictions. To support the experimental results, we conducted additional analysis using experimental data, shown in Fig.\ref{fig:inner_outer}. Here, $ \rho^{\text{est}}_{\text{1-LOCC}} $ ($ \rho^{\text{est}}_{\text{LO}} $) is the experimentally estimated quantum state using existing raw data from the LOCC (LO) protocol, which may not tomographically complete (see Methods). We then performed optimistic estimates in Eq.~(\ref{eq:lp_inner}) and rerun SDP.~(\ref{eq:sdp_outer}) for both inner approximation and outer approximation of the false negative error probability $ p_2 $ for the two quantum states under the LO measurement protocol.

As shown in Fig.~\ref{fig:inner_outer}, results for both quantum states are highly consistent under both approximations, suggesting the measurement strategy (Tables \ref{tab:px_mix} and \ref{Pylabel_ax}) effectively mitigates measurement-instability concerns. Most importantly, even under the optimistic estimate for LO, the experimental false-negative error probability $p_2$ for 1-LOCC remains lower, providing experimental evidence for the advantage of 1-LOCC. The remaining discrepancy from theoretical values is likely due to experimental imperfections.

\begin{table}[htbp!]
    \centering
    \caption{The $p_1$, $p_2^{LO}$, $p_2^{1-LOCC}$, and $G$ of theoretical, experimental and estimated states $\rho^*$, $\rho_{exp}$, $\rho_{est}^{LO}$, $\rho_{est}^{1-LOCC}$.}
    \fontsize{8}{11}\selectfont
    \setlength{\tabcolsep}{6pt}
    \begin{tabular}[t]{cccccc}
        \hline
        -&$p_1$&$p_2^{1-LOCC}$&$p_2^{LO}$&$G$\\
        \hline
        $\rho^*$&0.7481&0.0283&0.0944&0.0661\\
        $\rho_{exp}$&0.7481&0.0779&0.1390&0.0611\\
        $\rho_{est}^{LO}$&0.7481&0.0827&0.1402&0.0575\\
        $\rho_{est}^{1-LOCC}$ &0.7481&0.0812& 0.1388&0.0576\\
        \hline
    \end{tabular}\label{Tab: data}
\end{table}
    
\begin{table}[ht]
\centering
\caption{Adjustable angles of Half-Wave Plates (HWPs) and Liquid Crystals (LCs) used in the state preparation setup.}
\fontsize{8}{10}\selectfont
\setlength{\tabcolsep}{6pt}
\begin{tabular}{llll}
    \hline
    \multirow{2}{*}{Stage 1} & \multicolumn{1}{c}{HWP 1}    & \multicolumn{2}{c}{14.1008}  \\\cline{2-4}
                             & \multicolumn{1}{c}{HWP 2}    & \multicolumn{2}{c}{22.4197}  \\\hline \hline
    \multirow{9}{*}{Stage 2} & \multicolumn{1}{c}{$\phi$ 1}     & \multicolumn{2}{c}{-133.4403} \\\cline{2-4}
                             & \multicolumn{1}{c}{$\phi$ 2}     & \multicolumn{2}{c}{-116.7898} \\\cline{2-4}
                             & \multicolumn{1}{c}{$\phi$ 3}     & \multicolumn{2}{c}{138.4193}  \\\cline{2-4}
                             & \multirow{2}{*}{$T_{21}^*$}  & \multicolumn{1}{c}{HWP} & 220.4253  \\\cline{3-4}
                             &                              & \multicolumn{1}{c}{LC}  &  -24.2479 \\\cline{2-4}
                             & \multirow{2}{*}{$T_{20}^*$}  & \multicolumn{1}{c}{HWP} &  -22.3150 \\\cline{3-4}
                             &                              & \multicolumn{1}{c}{LC}  & -4.2391   \\\cline{2-4}
                             & \multirow{2}{*}{$T_{10}^*$}  & \multicolumn{1}{c}{HWP} & -13.8507  \\\cline{3-4}
                             &                              & \multicolumn{1}{c}{LC}  & 100.6367  \\\hline
\end{tabular}\label{Tab:angles}
\end{table}

\begin{table}
\caption{The small distributions $P(x)^*$ $P(y|ax)^*$ for 1-LOCC protocols and $P(xy)^*$ for LO protocols.}
\begin{minipage}{.2\linewidth}
\fontsize{8}{10}\selectfont
\setlength{\tabcolsep}{1pt}
\centering
\begin{tabular}{ *{2}{c} }
  \toprule
  $x$ & $P(x)$ \\
    \midrule
     1 & 0.2020\\
     2 & 0.4768\\
     3 & 0.3212\\
   \bottomrule
\end{tabular}
\end{minipage}%
\begin{minipage}{.5\linewidth}
\fontsize{8}{10}\selectfont
\setlength{\tabcolsep}{1pt}
\centering
\begin{tabular}{ *{5}{c} }
  \toprule
  $a$ &  $x$ &  $P(1|ax)$ &  $P(2|ax)$& $P(3|ax)$ \\
  \midrule
     1 & 1 & 0.0000 & 0.0378 & 0.9622\\
     1 & 2 & 1.0000 & 0.0000 & 0.0000\\
     1 & 3 & 0.0000 & 0.0000 & 1.0000\\
     2 & 1 & 0.0000 & 1.0000 & 0.0000\\
     2 & 2 & 1.0000 & 0.0000 & 0.0000\\
     2 & 3 & 0.0000 & 0.0089 & 0.9911\\
     3 & 1 & 0.0000 & 0.0139 & 0.9861\\
     3 & 2 & 0.0000 & 1.0000 & 0.0000\\
     3 & 3 & 0.2526 & 0.4960 & 0.2514\\
  \bottomrule
\end{tabular}
\end{minipage}
\begin{minipage}{.2\linewidth}
\fontsize{8}{10}\selectfont
\setlength{\tabcolsep}{1pt}
\centering
\begin{tabular}{ *{3}{c} }
\toprule
 $x$ &  $y$ &  $P(xy)$\\\hline
 1 & 1 & 0.0000\\
 1 & 2 & 0.1065\\
 1 & 3 & 0.1290\\
 2 & 1 & 0.3870\\
 2 & 2 & 0.1672\\
 2 & 3 & 0.0000\\
 3 & 1 & 0.0000\\
 3 & 2 & 0.0000\\
 3 & 3 & 0.2103\\
 \bottomrule
\end{tabular}\label{Tab: decision}
\end{minipage}%
\end{table}

\begin{table}[ht]
\centering
\caption{The LO and 1-LOCC shuffled protocol $P_{m}(x)$}
\fontsize{8}{10}\selectfont
\setlength{\tabcolsep}{10pt}
\begin{tabular}{cccc}
\hline
 $x$ & 0 & 1 & 2 \\\hline
 $P(x)$ & 0.2188 & 0.5155 & 0.2657\\\hline
\end{tabular}\label{tab:px_mix}
\end{table}

\begin{table}[ht]
\centering
\caption{The LO and 1-LOCC shuffled protocol $P_{m}(y'|ax)$}
\fontsize{8}{10}\selectfont
\setlength{\tabcolsep}{2pt}
\begin{tabular}{*{8}{c}}
\hline
 $a$ &  $x$ & $P(0|ax)$ &  $P(1|ax)$ &  $P(2|ax)$ &  $P(3|ax)$ &  $P(4|ax)$ &  $P(5|ax)$\\\hline
 1 & 1 & 0.0000 & 0.2435 & 0.2947 & 0.0000 & 0.0175 & 0.4443\\
 1 & 2 & 0.3754 & 0.1622 & 0.0000 & 0.4624 & 0.0000 & 0.0000\\
 1 & 3 & 0.0000 & 0.0000 & 0.3956 & 0.0000 & 0.0000 & 0.6044\\
 2 & 1 & 0.0000 & 0.2435 & 0.2947 & 0.0000 & 0.4618 & 0.0000\\
 2 & 2 & 0.3754 & 0.1622 & 0.0000 & 0.4624 & 0.0000 & 0.0000\\
 2 & 3 & 0.0000 & 0.0000 & 0.3956 & 0.0000 & 0.0054 & 0.5990\\
 3 & 1 & 0.0000 & 0.2435 & 0.2947 & 0.0000 & 0.0064 & 0.4553\\
 3 & 2 & 0.3754 & 0.1622 & 0.0000 & 0.0000 & 0.4624 & 0.0000\\
 3 & 3 & 0.0000 & 0.0000 & 0.3956 & 0.1527 & 0.2998 & 0.1519\\\hline
\end{tabular}\label{Pylabel_ax}
\end{table}

\begin{table}[ht]
\centering
\caption{Angles of HWPs to perform Alice's measurement.}
\fontsize{8}{10}\selectfont
\setlength{\tabcolsep}{6pt}
\begin{tabular}[t]{lccc}
    \hline
     &Computational&Fourier&Gell-Mann\\
    \hline
    HWP3&$0^\circ$&$-22.5^\circ$&$-22.5^\circ$\\
    HWP4&$0^\circ$&$17.6^\circ$&$45^\circ$\\
    HWP5&$0^\circ$&$22.5^\circ$&$0^\circ$\\
    \hline
\end{tabular}
\label{Tab: angle_measure}
\end{table}

\begin{table}[ht]
\centering
\caption{Voltage of EOMs to perform Bob's measurement.}
\fontsize{8}{10}\selectfont
\setlength{\tabcolsep}{6pt}
\begin{tabular}[t]{lccc}
    \hline
     &Computational&Fourier&Gell-Mann\\
    \hline
    EOM1&offset&pulse&pulse\\
    EOM2&offset&pulse&offset\\
    EOM3&pulse&offset&pulse\\
    \hline
\end{tabular}\label{Tab: voltage}
\end{table}

\begin{table}[ht]
\centering
\caption{The optimal instructions of LO and 1-LOCC scenarios for $\rho^*$}
\fontsize{6}{7}\selectfont
\setlength{\tabcolsep}{1.5pt}
\begin{tabular}{|l|l|l|l|c|c|l|l|l|l|c|c|}
\hline
$x$&$y$&$a$&$b$&$P_{LO}(xyN|ab)$&$P_{1-LOCC}(xyN|ab)$&$x$&$y$&$a$&$b$&$P_{LO}(xyN|ab)$&$P_{1-LOCC}(xyN|ab)$\\\hline
1&1&1&1&0&0&2&2&2&2&0&0\\\hline
1&1&1&2&0&0&2&2&2&3&0&0\\\hline
1&1&1&3&0&0&2&2&3&1&0&0\\\hline
1&1&2&1&0&0&2&2&3&2&0.1138&0\\\hline
1&1&2&2&0&0&2&2&3&3&0.1672&0.4768\\\hline
1&1&2&3&0&0&2&3&1&1&0&0\\\hline
1&1&3&1&0&0&2&3&1&2&0&0\\\hline
1&1&3&2&0&0&2&3&1&3&0&0\\\hline
1&1&3&3&0&0&2&3&2&1&0&0\\\hline
1&2&1&1&0&0&2&3&2&2&0&0\\\hline
1&2&1&2&0.1065&0.0076&2&3&2&3&0&0\\\hline
1&2&1&3&0.1065&0.0076&2&3&3&1&0&0\\\hline
1&2&2&1&0.1065&0.202&2&3&3&2&0&0\\\hline
1&2&2&2&0&0&2&3&3&3&0&0\\\hline
1&2&2&3&0&0&3&1&1&1&0&0\\\hline
1&2&3&1&0.0522&0.0028&3&1&1&2&0&0\\\hline
1&2&3&2&0.0573&0.0028&3&1&1&3&0&0\\\hline
1&2&3&3&0.0442&0&3&1&2&1&0&0\\\hline
1&3&1&1&0&0&3&1&2&2&0&0\\\hline
1&3&1&2&0.129&0.1453&3&1&2&3&0&0\\\hline
1&3&1&3&0.129&0.1944&3&1&3&1&0&0.0811\\\hline
1&3&2&1&0.0053&0&3&1&3&2&0&0.0811\\\hline
1&3&2&2&0&0&3&1&3&3&0&0\\\hline
1&3&2&3&0.129&0&3&2&1&1&0&0\\\hline
1&3&3&1&0.129&0.1992&3&2&1&2&0&0\\\hline
1&3&3&2&0.129&0.1992&3&2&1&3&0&0\\\hline
1&3&3&3&0&0&3&2&2&1&0&0.0028\\\hline
2&1&1&1&0.387&0.4768&3&2&2&2&0&0.0028\\\hline
2&1&1&2&0&0&3&2&2&3&0&0\\\hline
2&1&1&3&0.1836&0.046&3&2&3&1&0&0\\\hline
2&1&2&1&0&0&3&2&3&2&0&0.1593\\\hline
2&1&2&2&0.387&0.4768&3&2&3&3&0&0\\\hline
2&1&2&3&0.1781&0.046&3&3&1&1&0&0\\\hline
2&1&3&1&0.0639&0&3&3&1&2&0.2103&0.2145\\\hline
2&1&3&2&0.0708&0&3&3&1&3&0.2103&0.3212\\\hline
2&1&3&3&0.0455&0&3&3&2&1&0.2103&0.3183\\\hline
2&2&1&1&0.0835&0&3&3&2&2&0.2103&0.3183\\\hline
2&2&1&2&0&0&3&3&2&3&0&0\\\hline
2&2&1&3&0&0&3&3&3&1&0.1576&0.0807\\\hline
2&2&2&1&0.0841&0&3&3&3&2&0&0\\\hline
-&-&-&-&-&-&3&3&3&3&0.2103&0.0807\\\hline
\end{tabular}\label{Tab: Pxyn_ab}
\end{table}

\begin{table}[ht]
\centering
\caption{The final decision distributions $P(N|abxy)^*$ for LO and 1-LOCC scenarios.}
\fontsize{6}{7}\selectfont
\setlength{\tabcolsep}{1.5pt}
\begin{tabular}{|l|l|l|l|c|c|l|l|l|l|c|c|}
\hline
$x$&$y$&$a$&$b$&$P_{LO}(N|abxy)$&$P_{1-LOCC}(N|abxy)$&$x$&$y$&$a$&$b$&$P_{LO}(N|abxy)$&$P_{1-LOCC}(N|abxy)$\\\hline
 1 & 1 & 1 & 1 & 0.0000 & 0.0336 & 2 & 2 & 2 & 2 & 0.0000 & 0.5667\\\hline
 1 & 1 & 1 & 2 & 0.0000 & 0.0369 & 2 & 2 & 2 & 3 & 0.0000 & 0.0487\\\hline
 1 & 1 & 1 & 3 & 0.5000 & 0.5030 & 2 & 2 & 3 & 1 & 0.0000 & 0.0000\\\hline
 1 & 1 & 2 & 1 & 0.8333 & 0.9833 & 2 & 2 & 3 & 2 & 0.6806 & 0.0000\\\hline
 1 & 1 & 2 & 2 & 0.8333 & 0.9365 & 2 & 2 & 3 & 3 & 1.0000 & 1.0000\\\hline
 1 & 1 & 2 & 3 & 0.5000 & 0.0001 & 2 & 3 & 1 & 1 & 0.0000 & 0.0394\\\hline
 1 & 1 & 3 & 1 & 0.6667 & 0.3303 & 2 & 3 & 1 & 2 & 0.4000 & 0.5009\\\hline
 1 & 1 & 3 & 2 & 0.3333 & 0.2849 & 2 & 3 & 1 & 3 & 0.4000 & 0.5330\\\hline
 1 & 1 & 3 & 3 & 0.5000 & 0.7040 & 2 & 3 & 2 & 1 & 0.0000 & 0.0796\\\hline
 1 & 2 & 1 & 1 & 0.0000 & 0.0000 & 2 & 3 & 2 & 2 & 0.4000 & 0.5025\\\hline
 1 & 2 & 1 & 2 & 1.0000 & 1.0000 & 2 & 3 & 2 & 3 & 0.8000 & 0.7731\\\hline
 1 & 2 & 1 & 3 & 1.0000 & 1.0000 & 2 & 3 & 3 & 1 & 0.2000 & 0.5488\\\hline
 1 & 2 & 2 & 1 & 1.0000 & 1.0000 & 2 & 3 & 3 & 2 & 0.4000 & 0.0073\\\hline
 1 & 2 & 2 & 2 & 0.0000 & 0.0000 & 2 & 3 & 3 & 3 & 0.8000 & 0.6488\\\hline
 1 & 2 & 2 & 3 & 0.0000 & 0.0000 & 3 & 1 & 1 & 1 & 0.0000 & 0.1790\\\hline
 1 & 2 & 3 & 1 & 0.4900 & 1.0000 & 3 & 1 & 1 & 2 & 0.0000 & 0.0202\\\hline
 1 & 2 & 3 & 2 & 0.5383 & 1.0000 & 3 & 1 & 1 & 3 & 0.5000 & 0.5044\\\hline
 1 & 2 & 3 & 3 & 0.4153 & 0.0000 & 3 & 1 & 2 & 1 & 0.7500 & 0.3299\\\hline
 1 & 3 & 1 & 1 & 0.0000 & 0.0000 & 3 & 1 & 2 & 2 & 0.5000 & 0.2843\\\hline
 1 & 3 & 1 & 2 & 1.0000 & 0.7473 & 3 & 1 & 2 & 3 & 0.5000 & 0.7027\\\hline
 1 & 3 & 1 & 3 & 1.0000 & 1.0000 & 3 & 1 & 3 & 1 & 0.5000 & 1.0000\\\hline
 1 & 3 & 2 & 1 & 0.0410 & 0.9388 & 3 & 1 & 3 & 2 & 0.7500 & 1.0000\\\hline
 1 & 3 & 2 & 2 & 0.0000 & 0.0002 & 3 & 1 & 3 & 3 & 0.5000 & 0.0000\\\hline
 1 & 3 & 2 & 3 & 1.0000 & 0.9825 & 3 & 2 & 1 & 1 & 0.0000 & 0.0204\\\hline
 1 & 3 & 3 & 1 & 1.0000 & 1.0000 & 3 & 2 & 1 & 2 & 0.0000 & 0.0050\\\hline
 1 & 3 & 3 & 2 & 1.0000 & 1.0000 & 3 & 2 & 1 & 3 & 0.9913 & 0.9736\\\hline
 1 & 3 & 3 & 3 & 0.0000 & 0.0000 & 3 & 2 & 2 & 1 & 0.4957 & 1.0000\\\hline
 2 & 1 & 1 & 1 & 1.0000 & 1.0000 & 3 & 2 & 2 & 2 & 0.4870 & 1.0000\\\hline
 2 & 1 & 1 & 2 & 0.0000 & 0.0000 & 3 & 2 & 2 & 3 & 0.5130 & 0.0000\\\hline
 2 & 1 & 1 & 3 & 0.4744 & 0.0965 & 3 & 2 & 3 & 1 & 0.0087 & 0.0000\\\hline
 2 & 1 & 2 & 1 & 0.0000 & 0.0000 & 3 & 2 & 3 & 2 & 0.9913 & 1.0000\\\hline
 2 & 1 & 2 & 2 & 1.0000 & 1.0000 & 3 & 2 & 3 & 3 & 0.0087 & 0.0000\\\hline
 2 & 1 & 2 & 3 & 0.4600 & 0.0965 & 3 & 3 & 1 & 1 & 0.0000 & 0.0000\\\hline
 2 & 1 & 3 & 1 & 0.1652 & 0.5890 & 3 & 3 & 1 & 2 & 1.0000 & 0.6677\\\hline
 2 & 1 & 3 & 2 & 0.1829 & 0.6521 & 3 & 3 & 1 & 3 & 1.0000 & 1.0000\\\hline
 2 & 1 & 3 & 3 & 0.1176 & 0.0066 & 3 & 3 & 2 & 1 & 1.0000 & 1.0000\\\hline
 2 & 2 & 1 & 1 & 0.4993 & 0.7151 & 3 & 3 & 2 & 2 & 1.0000 & 1.0000\\\hline
 2 & 2 & 1 & 2 & 0.0000 & 0.4355 & 3 & 3 & 2 & 3 & 0.0000 & 0.0000\\\hline
 2 & 2 & 1 & 3 & 0.0000 & 0.0180 & 3 & 3 & 3 & 1 & 0.7493 & 1.0000\\\hline
 2 & 2 & 2 & 1 & 0.5027 & 0.8276 & 3 & 3 & 3 & 2 & 0.0000 & 0.0000\\\hline
 - & - & - & - & - & - & 3 & 3 & 3 & 3 & 1.0000 & 1.0000\\\hline
\end{tabular}\label{Tab: PN_abxy}
\end{table}
\clearpage

\end{document}